\documentclass[submission, Proceedings]{SciPost}

\hypersetup{
    colorlinks,
    linkcolor={red!50!black},
    citecolor={blue!50!black},
    urlcolor={blue!80!black}
}

\usepackage[bitstream-charter]{mathdesign}
\def\beq{\begin{equation}}
\def\eeq{\end{equation}}
\def\bea{\begin{eqnarray}}
\def\eea{\end{eqnarray}}
\def\beqn{\begin{eqnarray}} \def\eeqn{\end{eqnarray}}

\DeclareSymbolFont{usualmathcal}{OMS}{cmsy}{m}{n}
\DeclareSymbolFontAlphabet{\mathcal}{usualmathcal}

\begin{document}

\begin{center}{\Large \textbf{
Geometry and causality for efficient multiloop representations\\
}}\end{center}

\begin{center}
German F. R. Sborlini\textsuperscript{1,2$\star$}
\end{center}

\begin{center}
{\bf 1} Deutsches Elektronen-Synchrotron DESY, Platanenallee 6, 15738 Zeuthen, Germany. \\
{\bf 2} Instituto de F\'{\i}sica Corpuscular, Universitat de Val\`{e}ncia -- Consejo Superior de Investigaciones Cient\'{\i}ficas, Parc Cient\'{\i}fic, E-46980 Paterna, Valencia, Spain.
\\
* german.sborlini@desy.de
\end{center}

\begin{center}
\today
\end{center}


\definecolor{palegray}{gray}{0.95}
\begin{center}
\colorbox{palegray}{
  \begin{tabular}{rr}
  \begin{minipage}{0.1\textwidth}
    \includegraphics[width=35mm]{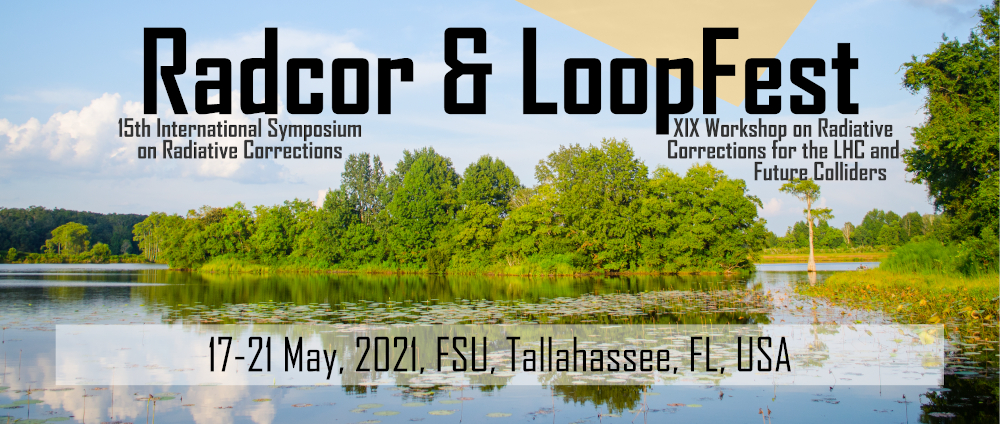}
  \end{minipage}
  &
  \begin{minipage}{0.85\textwidth}
    \begin{center}
    {\it 15th International Symposium on Radiative Corrections: \\Applications of Quantum Field Theory to Phenomenology,}\\
    {\it Tallahasse, FL, USA, 17-21 May 2021} \\
    \doi{10.21468/SciPostPhysProc.?}\\
    \end{center}
  \end{minipage}
\end{tabular}
}
\end{center}

\section*{Abstract}
{\bf
Multi-loop scattering amplitudes constitute a serious bottleneck in current high-energy physics computations. Obtaining new integrand level representations with smooth behaviour is crucial for solving this issue, and surpassing the precision frontier. In this talk, we describe a new technology to rewrite multi-loop Feynman integrands in such a way that non-physical singularities are avoided. The method is inspired by the Loop-Tree Duality (LTD) theorem, and uses geometrical concepts to derive the causal structure of any multi-loop multi-leg scattering amplitude. This representation makes the integrand much more stable, allowing faster numerical simulations, and opens the path for novel re-interpretations of higher-order corrections in QFT.}

\vspace{10pt}
\noindent\rule{\textwidth}{1pt}
\tableofcontents\thispagestyle{fancy}
\noindent\rule{\textwidth}{1pt}
\vspace{10pt}

\section{Introduction}
\label{sec:intro}
Higher-order contributions involve dealing with vacuum quantum fluctuations which are encoded through complicated multi-loop multi-leg Feynman diagrams. They currently constitute a bottleneck to break the precision frontier. Many techniques have been developed to calculate these objects \cite{Heinrich:2020ybq}, including analytic and numeric approaches. The presence of singularities forces the introduction of regularization prescriptions, such as Dimensional Regularization (DREG) \cite{Ashmore:1972uj,Cicuta:1972jf,tHooft:1972tcz}, which prevents a straightforward numerical implementation. Several regularization methods are available in the market \cite{Gnendiger:2017pys,TorresBobadilla:2020ekr}, but still DREG is considered as the default one for computing higher-order corrections for colliders.

Once the singularities are regularized, they can be removed and the resulting expression can be numerically computed. The cancellation of singularities is achieved by the inclusion of ultraviolet (UV) and infrared (IR) counter-terms. Whilst the first ones are obtained through the renormalization of the theory, the IR ones can be generated by studying the real emission contributions, as well as the so-called initial state radiation (ISR). In any case, the traditional framework involves a separate calculation of the ingredients, namely:
\begin{enumerate}
    \item The virtual contributions, i.e. the diagrams containing loops defined over a Minkowski integration space.
    \item The real correction, determined by diagrams with extra real-radiation integrated over the phase-space (i.e. an Euclidean integration space).
    \item Ultraviolet and infrared counter-terms, which are proportional to the lower orders and can be easily integrated analytically.
\end{enumerate}
The singular structures of the different contributions perfectly match, and the sum of the \emph{integrated} terms is free of divergences. So, in the standard framework, the cancellation of singularities takes place \emph{after integration}.

With the purpose of simplifying the numerical implementation, we have been developing a strategy that aims to achieve the cancellation of singularities \emph{before integration}: this is the so-called \emph{local} cancellation. This approach is based on the Loop-Tree Duality (LTD) \cite{Catani:2008xa,Rodrigo:2008fp,Bierenbaum:2010cy,Bierenbaum:2012th,Buchta:2014dfa,Rodrigo:2016hqc,deJesusAguilera-Verdugo:2021mvg} to re-write the loop amplitudes as dual contributions defined in an Euclidean space. In this way, after introducing proper mappings, it is possible to relate the kinematics of the real and dual components, leading to a unified description and a natural integrand-level cancellation of IR singularities \cite{Hernandez-Pinto:2015ysa,Sborlini:2015uia,Sborlini:2016fcj,Sborlini:2016gbr,Sborlini:2016hat}. Moreover, the ISR and UV counter-terms can be also expressed as phase-space integrals to locally cancel the corresponding singularities \cite{Driencourt-Mangin:2017gop,Driencourt-Mangin:2019aix,Driencourt-Mangin:2019yhu,Plenter:2020lop,Prisco:2020kyb}. This constitutes the Four-Dimensional Unsubtraction (FDU) approach, which allows to by-pass DREG and provides a purely four-dimensional representation of physical observables that is totally free of IR and UV singularities.

Regarding the treatment of the virtual contributions, we have recently shown that the LTD framework leads to a manifestly causal representation of multi-loop multi-leg scattering amplitudes \cite{Aguilera-Verdugo:2019kbz,Verdugo:2020kzh,Ramirez-Uribe:2020hes,MANIFESTLYCAUSAL,Aguilera-Verdugo:2020kzc}. The causal structure of scattering amplitudes has been extensively studied \cite{Cutkosky:1960sp,Tomboulis:2017rvd,Runkel:2019yrs,Runkel:2019zbm,Capatti:2020ytd} since it leads to noticeable simplifications and the cancellation of spurious unphysical singularities. Even if these singularities are integrable, they introduce serious numerical instabilities: in consequence, causal representations turn out to be more suitable for efficient numerical calculations. Also, causality provides a powerful tool to reconstruct scattering amplitudes by using information about the physical threshold singularities and discontinuities \cite{Cahill:1973qp,Benincasa:2020aoj}.

In this article, we comment on very recent developments concerning the description of the causal structures inside multi-loop multi-leg scattering amplitudes. Using a geometrical construction inspired by graph theory, we establish a set of rules that allows to compute all the possible entangled thresholds contributing to the causal representation \cite{Sborlini:2021owe}. We rely on the identification of binary connected partitions of diagrams (which strictly corresponds to the so-called causal propagators) and compatible momenta orientations (usually called causal fluxes) \cite{Ramirez-Uribe:2021ubp}. This geometrical approach is closely related to the all-order representations presented in Refs. \cite{Bobadilla:2021pvr,TorresBobadilla:2021ivx} that directly reproduce the terms originated by the explicit nested residue calculation \cite{Verdugo:2020kzh,Aguilera-Verdugo:2020kzc,Ramirez-Uribe:2020hes,JesusAguilera-Verdugo:2020fsn}.


\section{Brief introduction to the Loop-Tree Duality}
\label{sec:LTDintro}
Let us consider a generic $L$-loop $N$-point scattering amplitude. We group the momenta of the propagators according to their dependence on the integration variables, i.e. the primitive loop momenta $\{\ell_i\}_{i=1,\ldots,L}$. Given a set $s$, a generic propagator $i \in s$ has momentum given by $q_{i}=\sum_j \, \beta_j^{s} \ell_j \, + \, k_{i}$, with $\beta_j^{s} \in\{-1,0,1\}$ and $k_{i}$ a linear combination of external momenta $\{p_r\}_{r=1,\ldots,N}$. With this convention, the scattering amplitude can be represented as 
\beqn
\mathcal{A}_{N}^{\left(L\right)}&=&\int_{\ell_{1},\hdots,\ell_{L}}\, \sum_j\, \mathcal{N}_j\left(\left\{ \ell_{i}\right\} _{L},\left\{ p_{j}\right\} _{N}\right) \, \times \, G_{F}\left(1,\hdots,n\right)\,,
\label{eq:AmplitudGeneral}
\eeqn
with $G_{F}\left(1,\ldots,n\right)=\prod_{i\in1\cup\cdots\cup n}\,(G_{F}(q_{i}))^{\alpha_{i}}$, the product of Feynman propagators associated to the momenta set $\{1,\ldots,n\}$. It is useful to write
\beq  
G_F(q_i)=\frac{1}{q_{i,0}^2-(q_{i,0}^{(+)})^2} \, ,
\eeq
where $q_{i,0}^{(+)}=\sqrt{\vec{q_i}^2+m_i^2-\imath 0}$ is the positive on-shell energy associated to $q_i$.

As carefully explained in Ref. \cite{Verdugo:2020kzh}, the LTD representation of Eq. (\ref{eq:AmplitudGeneral}) is obtained by removing one d.o.f. per loop, making use of a recursive computation of nested residues. Explicitly, if $d{\cal A}_{N}^{\left(L\right)}(1,\ldots,n)$ denotes the integrand of Eq. (\ref{eq:AmplitudGeneral}), the first application of Cauchy's residue theorem leads to
\beqn
d{\cal A}_D^{(L)}(1;\ldots,n) &=& \sum_{i \in 1} {\rm Res}\left(d{\cal A}_{N}^{\left(L\right)}, {\rm Im}(q_{i,0})<0 \right) \, ,
\label{eq:FirstSTEP}
\eeqn
where the semicolon is introduced to separate the sets with on-shell lines (left) from those that remain off-shell (right). After the $r$-th iteration, we have
\beqn
d{\cal A}_D^{(L)}(1,\ldots,r;r+1,\ldots,n) &=& \sum_{i \in r} {\rm Res}\left(d{\cal A}_D^{\left(L\right)}{(1,\ldots,r-1;r,\ldots,n)}, {\rm Im}(q_{i,0})<0 \right) \, .
\label{eq:IterationSTEP}
\eeqn
In order to obtain the LTD dual representation, we need to iterate the procedure $L$ times. After all the loops have been opened to trees by setting on-shell $L$ lines, the dual expression is written in terms of positive on-shell energies and products of external momenta.

Even if the technology of nested residues leads to explicit recursive relations and compact formulae \cite{Aguilera-Verdugo:2020kzc,Ramirez-Uribe:2020hes,JesusAguilera-Verdugo:2020fsn}, further simplifications take place when adding all the dual contributions together. In that case, the final result only involves same-sign combinations of on-shell energies, and we claim that \cite{TorresBobadilla:2021ivx,Sborlini:2021owe,Bobadilla:2021pvr}
\beqn
 {\cal A}_{N}^{\left(L\right)}&=&(-1)^k \,\sum_{\sigma \in \Sigma}  \, \int_{\vec{\ell}_{1},\cdots,\vec{\ell}_{L}}\frac{{\cal N}_{\sigma}(\{q_{r,0}^{(+)}\},\{p_{j,0}\})}{x_{n}} \, \times \prod_{i=1}^{k} \frac{1}{\lambda_{\sigma(i)}}\, + \, (\sigma \leftrightarrow \bar{\sigma})  \, ,
\label{eq:MasterFormula}
\eeqn
fully describes the causal structure of any multi-loop multi-leg Feynman diagram or collection of Feynman diagrams. The causal propagators, i.e.
\beq
\lambda_j^\pm  \equiv \sum_{i \in X_j} q_{i,0}^{(+)} \pm k_j \, ,
\eeq
codify the physical thresholds of the amplitude, whilst all the possible causal entangled thresholds are inside the set $\Sigma$. Each combination in $\Sigma$ contains products of $k$ causal propagators, being $k$ the order of the diagram. It is possible to prove that the order is given by $k=V-1$ and equals the number of off-shell lines in each allowed dual cut \cite{Verdugo:2020kzh,Sborlini:2021owe}. 


\section{Geometry of loop amplitudes}
\label{sec:Geometry}
Within perturbation theory, scattering amplitudes are described in terms of Feynman diagrams. These diagrams are graphs built from edges and vertices: edges represent virtual particles propagating between interaction vertices. As explained in Refs. \cite{JesusAguilera-Verdugo:2020fsn,TorresBobadilla:2021ivx,Sborlini:2021owe}, when considering the dual representation of any multi-loop multi-leg Feynman diagram, edges connecting the same pair of vertices are equivalent to a single one. This new \emph{multi-edge} is described by
\beq
q_{G,0} = \sum_{i \in G} \, q_{i,0} \quad , \quad q_{G,0}^{(+)} =   \sum_{i \in G} \, q_{i,0}^{(+)} \, ,
\eeq
where $G=\{i_1,\ldots,i_g\}$ are the lines being merged. By replacing edges by multi-edges, we obtain the so-called reduced Feynman graph, which contain all the required information to unveil the causal structure of the original diagram.

Once the reduced Feynman graph is obtained, we need to define a basis composed by the multi-edges and the external momenta, namely $Q=\{Q_1,\ldots,Q_M;p_1,\ldots,p_{N-1}\}$. Global momentum conservation is implicitly used here by writing $p_N = - \sum p_i$. At this point, we can represent each interaction vertex using their coordinates in the basis $Q$: outgoing (incoming) edges are positive (negative). Then, we can built the so-called \emph{vertex matrix}, ${\cal V}$, whose rows are the representation of the vertices in the basis $Q$. This is the fundamental object in our formalism, since we can extract all the required information by implementing operations on this matrix. In the following, we will explain how to obtain all the possible causal propagators and how to detect the allowed entangled causal configurations.


\subsection{Causal propagators}
\label{ssec:CausalPropagators}
Given a reduced Feynman graph, let us consider the set of all the vertices, $V=\{1,2,3,\ldots\}$. Then, consider all the possible ways to split the graph into two parts, which is equivalent to take binary partitions of $V$, i.e. ${\cal P}_V = \{ \{1\}, \{2\} , \ldots , \{1,2\}, \{1,3\}, \ldots\}$. In the definition of ${\cal P}_V$ we only include the sets with the minimal number of vertices, since we can define the equivalence relation $r \equiv r^c$ where $r^c=V/r$. Also, given a partition $p$, we say that it is \emph{connected} if for any pair of vertices $\{v_i,v_j\} \in p$, there exists a path (i.e. a sequence of multi-edges) contained in $p$ from $v_i$ to $v_j$. With this in mind, we define the set of \emph{connected binary partitions}, ${\cal P}_V^C$, as the subset of elements of $p \in {\cal P}_V$ such that $p$ and $p^c$ are connected. 

It turns out that there is a correspondence between the elements of ${\cal P}_V^C$ and the causal propagators. Given a connected binary partition, we have a set of vertices $p=\{v_{p_1},\ldots,v_{p_r}\}$ and each of them is associated to a unique conservation equation (i.e. the momentum conservation imposed by Feynman rules). We define the \emph{conjugated causal propagator}, $\bar{\lambda}_{p}$, as the sum of the energy components of the momenta involved in the vertices $\{v_{p_1},\ldots,v_{p_r}\}$. Finally, we can generate the causal propagators starting from $\bar{\lambda}_{p}$ and applying the transformation
\beq
\bar{\lambda}_p \to \lambda^{\pm}_p = \sum_j \beta_j \, Q^{(+)}_{j,0} \pm \sum_{i=1}^{N-1} \gamma_i \,  p_{i,0} \, ,
\label{eq:Transformation}
\eeq  
where $\beta_j, \gamma_i \in \{1,0\}$. Notice that this definition involves the positive on-shell energies of the corresponding multi-edges crossing the partition $p$. Also, it reproduces the alignment of their on-shell modes, which codifies the causal flux of energy through the partition.


\subsection{Selection rules}
\label{ssec:Selection}
Each causal propagator is associated to one threshold singularity in the original Feynman diagram. Given a Feynman diagram of order $k = V - 1$, we know that its causal representation involves products of $k$ causal propagators \cite{Verdugo:2020kzh,Bobadilla:2021pvr,Sborlini:2021owe}. In other words, each term in the causal representation is associated to a compatible entanglement of $k$ causal thresholds. To detect all the possible combinations of $k$ compatible \emph{causal entangled thresholds}, we present the following selection criteria \cite{Sborlini:2021owe}:
\begin{enumerate}
\setcounter{enumi}{0}
    \item \emph{All the lines are crossed}: Since entangled thresholds are originated from the superposition of cuts, and they describe a decomposition of the diagram in tree-level-like objects, all the multi-edges must be cut.
    \item \emph{Absence of threshold intersections}: Two compatible causal propagators $\lambda_p$ and $\lambda_q$ must fulfill that the associated connected sets of vertices are disjoint or totally included in the biggest set.
    \item \emph{Compatible causal flow of the multi-edges}: If all the multi-edges contributing to the entangled cuts appear with the same orientation, then they are compatible. This is equivalent to have an acyclic directed graph dressed with a subset of causal propagators $\{\lambda_{i_1},\ldots,\lambda_{i_k}\}$ \cite{Ramirez-Uribe:2021ubp}.
    \item \emph{Causal propagator orientation}: Once criteria 1-3 were applied, it is necessary to determine the sign of the transformation in Eq. (\ref{eq:Transformation}). If the external momenta and the oriented multi-edges are both outgoing, then $\bar{\lambda}_p \to \lambda_p^+$. Otherwise, $\bar{\lambda}_p \to \lambda_p^-$.
\end{enumerate}
The application of criteria 1-4 determines the set of all the allowed causal entangled thresholds, $\bar{\Sigma}$. Going back to Eq. (\ref{eq:MasterFormula}) for scalar amplitudes (i.e. ${\cal N}=1$), we can use $\Sigma = \bar{\Sigma}$ by introducing symmetry factors. These symmetry factors account for a degeneration due to global momentum conservation. In order to break the degeneration, we need to introduce a fifth selection criteria, which implies adding extra multi-edges composed by external particles and applying criteria 1-4 \cite{Sborlini:2021owe}. It is worth appreciating that the application of criterion 3 implies consistently ordering $2^M$ multi-edges. Since the computational complexity scales exponentially with the number of multi-edges, this could be a potential bottleneck for multi-leg multi-loop amplitudes. For this reason, novel strategies based on quantum algorithms are starting to be explored \cite{Ramirez-Uribe:2021ubp,Sborlini:2021PREPARATION}.

\section{Example: causal representation of a pentagon}
\label{sec:Pentagon}
\begin{figure}[ht]
\begin{center}
\includegraphics[width=0.50\textwidth]{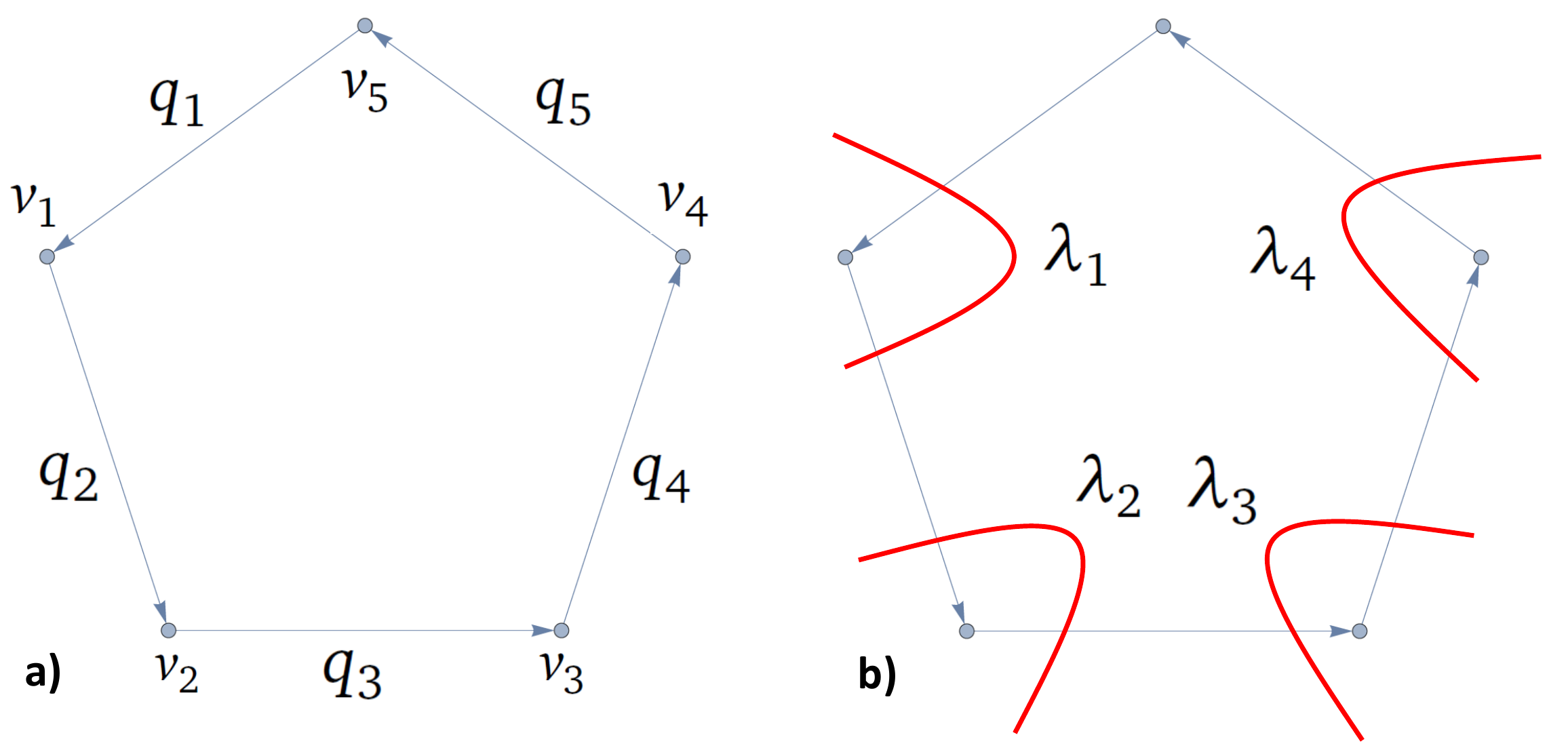} 
\caption{\textbf{a)} Feynman graph representing a pentagon, with the multi-edge and vertex labelling used in the analysis presented in the text. \textbf{b)} Example of an allowed causal entangled threshold contributing to the LTD causal representation.
\label{fig:1}}
\end{center}
\end{figure}

As a practical example, we consider a scalar five-point one-loop function, i.e the pentagon shown in Fig. \ref{fig:1}. After constructing the vertex matrix, we proceed to detect all the possible binary connected partitions. We obtain 10 conjugated causal propagators, i.e.
\beq
{\cal P}_V^{C} = \{\{1\}, \{2\},\{3\},\{4\},\{5\},\{1,2\},\{1,5\},\{2,3\},\{3,4\},\{4,5\} \} \, ,
\label{eq:ParticionesBinariasConectadas}
\eeq
described in terms of the vertices involved in the partition. By using the transformation defined in Eq. (\ref{eq:Transformation}), we can explicitly write the causal propagators as
\beqn  
\nonumber && \lambda_j^\pm = q_{j,0}^{(+)} + q_{j+1,0}^{(+)} \pm p_{j,0} \quad {\rm for} \quad j=\{1,2,3,4\} \, ,
\\ \nonumber && \lambda_5^\pm = q_{5,0}^{(+)} + q_{1,0}^{(+)} \pm ( p_{1,0} + p_{2,0}+ p_{3,0}+p_{4,0}) \, ,
\\ && \nonumber \lambda_6^\pm = q_{1,0}^{(+)} + q_{3,0}^{(+)} \pm ( p_{1,0} + p_{2,0}) \, , \quad \lambda_7^\pm = q_{2,0}^{(+)} + q_{5,0}^{(+)} \pm ( p_{2,0}+ p_{3,0}+p_{4,0}) \, ,
\\ \nonumber && \lambda_8^\pm = q_{2,0}^{(+)} + q_{4,0}^{(+)} \pm (  p_{2,0}+ p_{3,0}) \, , \quad \lambda_9^\pm = q_{3,0}^{(+)} + q_{5,0}^{(+)} \pm ( p_{3,0}+p_{4,0}) \, ,
\\ && \lambda_{10}^\pm = q_{1,0}^{(+)} + q_{4,0}^{(+)} \pm ( p_{1,0} + p_{2,0}+ p_{3,0}) \, .
\label{eq:Lambdas}
\eeqn 
Once all the causal propagators are known, we apply the criteria 1-4 explained in Sec. \ref{ssec:Selection} to select the allowed causal entangled thresholds. Since we have 10 causal propagators and the order of the diagram is $k=4$, there are 210 combinations. However, criteria 1-2 reduce the possibilities to only 60. Imposing criteria 3-4, we obtain the following 55 combinations
\beqn 
\nonumber \bar{\Sigma} &=& \{ \{1, 2, 3, 4\}, \{1, 2, 3, 5\}, \{1, 2, 3, 7\}, \{1, 2, 3, 9\}, \{1, 2, 4, 5\}, \{1, 2, 4, 7\}, \{1, 2, 4, 8\}, 
\\ \nonumber && \{1, 2, 4, 9\}, \{1, 2, 4, 10\}, \{1, 2, 5, 8\}, \{1, 2, 5, 10\}, \{1, 2, 7, 8\}, \{1, 3, 4, 5\}, \{1, 3, 4, 6\},
\\ \nonumber && \{1, 3, 4, 7\}, \{1, 3, 4, 8\}, \{1, 3, 4, 10\}, \{1, 3, 5, 6\}, \{1, 3, 5, 8\}, \{1, 3, 5, 9\}, \{1, 3, 5, 10\}, 
\\ \nonumber && \{1, 3, 6, 9\}, \{1, 3, 7, 8\}, \{1, 3, 7, 9\}, \{1, 4, 5, 6\}, \{1, 4, 5, 9\}, \{1, 4, 6, 9\}, \{1, 4, 6, 10\}, 
\\ \nonumber && \{1, 4, 7, 9\}, \{1, 5, 6, 10\}, \{2, 3, 4, 5\}, \{2, 3, 4, 6\}, \{2, 3, 4, 10\}, \{2, 3, 5, 6\}, \{2, 3, 5, 7\}, 
\\ \nonumber && \{2, 3, 5, 9\}, \{2, 3, 5, 10\}, \{2, 3, 6, 9\}, \{2, 4, 5, 6\}, \{2, 4, 5, 7\}, \{2, 4, 5, 8\}, \{2, 4, 5, 9\}, 
\\ \nonumber && \{2, 4, 6, 9\}, \{2, 4, 6, 10\}, \{2, 4, 8, 10\}, \{2, 5, 6, 10\}, \{2, 5, 7, 8\}, \{2, 5, 8, 10\}, \{3, 4, 5, 7\},
\\ && \{3, 4, 5, 8\}, \{3, 4, 8, 10\}, \{3, 5, 7, 8\}, \{3, 5, 7, 9\}, \{3, 5, 8, 10\}, \{4, 5, 7, 9\}\} \, ,
\label{eq:SetEntangledCompleto}
\eeqn
where we use the short-hand notation $i\equiv \lambda_i$. Notice that the signature of each entangled causal threshold, i.e. whether we should put $\lambda_i^+$ or $\lambda_i^-$, is unambiguously by criterion 4.

Since the pentagon corresponds to a non-maximally connected topology, some elements of $\bar{\Sigma}$ might be degenerated due to global momentum conservation. In fact, applying the criterion 5 as defined in Ref. \cite{Sborlini:2021owe} for the one-loop case, we obtain the reduced (minimal) set
\beqn 
\Sigma &=& \{\{1, 2, 3, 4\}, \{1, 2, 3, 9\}, \{1, 2, 4, 7\}, \{1, 2, 4, 8\}, \{1, 2, 4, 9\}, \{1, 2, 7, 8\}, \{1, 3, 4, 6\}, 
\\ \nonumber && \{1, 3, 4, 8\}, \{1, 3, 4, 10\}, \{1, 3, 5, 9\}, \{1, 3, 5, 10\}, \{1, 3, 6, 9\}, \{1, 3, 7, 8\}, \{1, 3, 7, 9\}, 
\\ \nonumber && \{1, 4, 5, 6\}, \{1, 4, 5, 9\}, \{1, 4, 6, 9\}, \{1, 4, 6, 10\}, \{1, 4, 7, 9\}, \{1, 5, 6, 10\}, \{2, 3, 4, 6\}, 
\\ \nonumber && \{2, 3, 6, 9\}, \{2, 4, 5, 6\}, \{2, 4, 5, 7\}, \{2, 4, 6, 9\}, \{2, 4, 6, 10\}, \{2, 4, 8, 10\}, \{2, 5, 6, 10\}, 
\\  \nonumber && \{2, 5, 7, 8\}, \{2, 5, 8, 10\}, \{3, 4, 8, 10\}, \{3, 5, 7, 8\}, \{3, 5, 7, 9\}, \{3, 5, 8, 10\}, \{4, 5, 7, 9\}\} ,
\label{eq:SetNODegenerado}
\eeqn
which is composed by 35 elements and directly leads to a causal representation as defined in Eq. (\ref{eq:MasterFormula}) with ${\cal N}_\sigma =1$ for every $\sigma \in \Sigma$. In particular, it is worth noticing that the configurations
\beq 
 \{\{1,2,3,4\},\{1,2,3,5\},\{1,2,4,5\},\{1,3,4,5\},\{2,3,4,5\} \}  \, ,
 \label{eq:Degenerados}
\eeq
are degenerated. They can be graphically interpreted as rotations of the causal entangled threshold shown in Fig. \ref{fig:1} (right). Since we impose momentum conservation in the external legs and replace $p_5$ in terms of the other momenta, we implicitly \emph{marked} $v_5$ as a distinguished vertex. The algorithm that we defined for criterion 5 uses this information, and this is the reason why $\{1,2,3,4\}$ is kept in $\Sigma$ whilst the others in Eq. (\ref{eq:Degenerados}) are eliminated.

\section{Conclusions}
\label{sec:conclusions}
We describe a purely geometrical formalism to obtain a causal dual representation of any multi-loop multi-leg Feynman diagram. By generating all the possible connected binary partitions of a reduced Feynman graph, we manage to obtain the set of causal propagators which codify the physical thresholds that are present in the original diagram. Then, by applying a set of geometrical criteria, we identify all the possible causal entangled thresholds, $\bar{\Sigma}$. This set contains information about all the thresholds that might occur simultaneously.

It turns out that it is possible to conjecture a general formula to describe a manifestly causal dual representation for any multi-loop multi-leg scattering amplitude. The master equation presented in Eq. (\ref{eq:MasterFormula}) is supported by the findings reported in Refs. \cite{Bobadilla:2021pvr,Sborlini:2021owe,TorresBobadilla:2021ivx}. In this article, we present a concrete application that shows how to obtain a causal representation for the pentagon, and compare the resulting structure with the conjectured formula. Deeper studies are required to properly understand the degeneration introduced because of global momentum conservation. This might hide important properties regarding the underlying structures of scattering amplitudes and, ultimately, physical cross-sections in QFT.

\section*{Acknowledgments}
I am grateful to Germ\'an Rodrigo for fruitful discussions and IFIC for the hospitality during the realization of this article. Also, I thank J. Aguilera-Verdugo, R. Hern\'andez-Pinto, S. Ram\'irez-Uribe and A. Renter\'ia-Olivo for reading and commenting the manuscript before publication.

\paragraph{Funding information}
This article is based upon work from COST Action PARTICLEFACE CA16201, supported by COST (European Cooperation in Science and Technology). \href{https://www.cost.eu/}{www.cost.eu}.


\nolinenumbers

\end{document}